\documentclass[aps,prb,twocolumn,groupedaddress]{revtex4-1}
\usepackage{graphicx}
\usepackage{amsmath}

\begin{document}


\title{Two-component uniform spin susceptibility in superconducting HgBa$_{2}$CuO$_{4+\delta}$ single crystals determined with $^{63}$Cu and $^{199}$Hg NMR}



\author{J\"urgen Haase$^{1}$}
\email[]{j.haase@physik.uni-leipzig.de} 
\author{Damian Rybicki$^1$} 
\author{Charles P. Slichter$^2$}

\author{Martin~Greven$^3$}
\author{Guichuan Yu$^3$}
\author{Yuan Li$^4$}
\author{Xudong Zhao$^{3,5}$}
\affiliation{$^1$ Faculty of Physics and Earth Science, University of Leipzig, Linn$\acute{e}$stra\ss e 5, 04103 Leipzig, Germany}
\affiliation{$^2$ Department of Physics, University of Illinois at Urbana-Champaign, Urbana IL 61801, USA}
\affiliation{$^3$ School of Physics and Astronomy, University of Minnesota, Minneapolis, Minnesota 55455, USA}
\affiliation{$^4$Max-Planck-Institut f\"ur Festk\"orperforschung, Heisenbergstrasse 1, D-70569 Stuttgart, Germany}
 \affiliation{$^5$College of Chemistry, Jilin University, Changchun 130012, China}

\date{\today}

\begin{abstract}
$^{63}$Cu and $^{199}$Hg NMR shifts for an optimally doped and underdoped HgBa$_{2}$CuO$_{4+\delta}$ single crystal are reported, and the temperature dependence dictates a two-component description of the uniform spin susceptibility. The first component, associated with the pseudogap phenomenon in the NMR shifts, decreases already at room temperature and continues to drop as the temperature is lowered, without a drastic change at the transition temperature into the superconducting state. The second component is temperature independent above the superconducting transition temperature and vanishes rapidly below it. It is a substantial part of the total $T$ dependent susceptibility measured at both nuclei. 
\end{abstract}



\maketitle


\section{Introduction}
One of the most challenging questions in condensed matter theory is embodied in the physics of high-temperature superconducting cuprates and concerns their transition from a Mott antiferromagnetic insulating state into a conductor and superconductor by removing or adding electrons, i.e., by doping them. Clearly, understanding the electronic spin or magnetic spin susceptibility $ \chi_{\rm S}$ of these materials is of great interest. In particular, it is important to know whether the uniform spin susceptibility can be understood in terms of a single electronic spin component that changes with doping and temperature across the phase diagram. Nuclear magnetic resonance (NMR) measures $ \chi_{\rm S}$  locally through the hyperfine coupling that causes, e.g., a relative shift $K_{{\rm S},\eta} = {\rm p}_\eta \cdot \chi_{\rm S}$ of the nuclear resonance frequency, where p$_\eta$ is the anisotropic hyperfine coupling constant \cite{Knight1949} and $\eta$ denotes the direction of the magnetic field with respect to the crystal axes. With NMR as a bulk probe, such measurements are reliable since they are insensitive to dilute impurities or magnetic inclusions. 

If all nuclei, $n$, of a given material experience the same normalized, proportional shift change ${\Delta {^nK}_{{\rm S},{\eta}}/{{^n{\rm p}}_{\eta}}=\Delta \chi_{\rm S}}$ in the temperature interval $\Delta T$, this is evidence for a single-component response. On the other hand, if one finds experimentally that this shift change with temperature differs for different nuclei, this proves the failure of a single-component view. This important question was addressed with NMR measurements early on. And it was concluded, rather quickly, with experiments on YBa$_{\rm 2}$Cu$_{\rm 3}$O$_{\rm 6.63}$ \cite{Takigawa1991} and later YBa$_{\rm 2}$Cu$_{\rm 4}$O$_{\rm 8}$ \cite{Bankay1994} that a single-component description of the cuprates is  appropriate. However, pioneering measurements of the total susceptibility indicated two-component behavior of different cuprates \cite{Johnston1989}, and NMR relaxation studies did not agree with a single-component model \cite{Walstedt1994}.

Recently, some of us showed with extensive NMR shift measurements that La$_{1.85}$Sr$_{0.15}$CuO$_4$ does \textit{not} show a single-component spin response \cite{Haase2009}, and since high quality HgBa$_{2}$CuO$_{4+\delta}$ single crystals had become available \cite{Zhao2006, Barisic2008}, we address here the important question whether the two-component behavior is a mere peculiarity of La$_{1.85}$Sr$_{0.15}$CuO$_4$ or a more universal property of the cuprates.

We have acquired a comprehensive set of NMR data for an underdoped and optimally doped sample\cite{Rybicki2011a}, and we will show that both materials require a two-component description.
Concurrent with the present work, some of us were involved with developing a new anvil cell NMR probe \cite{Haase2009a} for sensitive, high-pressure NMR experiments to enable, e.g., the investigation of the cuprates under pressures that had not be done before. Recent success was reported with $^{17}$O NMR of YBa$_{\rm 2}$Cu$_{\rm 4}$O$_{\rm 8}$ at pressures up to 63 kbar \cite{Meissner2011}. It was found, based in part on the results presented here, that the spin susceptibility in this system also requires a two-component description.

\section{Experimental}
The two crystals\cite{Zhao2006} used in this study were annealed to result in one optimally doped crystal with $T_{\rm c}$ = 97 K (mass 30.3 mg) that we label X97 and one underdoped crystal with $T_{\rm c} $ = 74 K (mass 3.3 mg) that we label X74\cite{Barisic2008}.
$^{63}$Cu and $^{199}$Hg NMR shifts as a function of temperature $T$ were recorded in a magnetic field B$_0=11.75$ T for two different orientations of the tetragonal single crystals with respect to the external magnetic field: with the field parallel ($c\parallel B_0$) and perpendicular ($c \bot B_0$) to the crystal $c$-axis. NMR shifts were referenced using a fine metallic copper powder. For $^{63}$Cu we used the published value of the $^{63}$Cu Knight shift of 3820 ppm.\cite{Lutz1978} $^{199}$Hg shifts were referenced to (CH$_3$)$_2$Hg using the referencing procedure described by Harris\cite{Harris2008}.  Below $T_{\rm c}$, in particular for $^{199}$Hg, the sensitivity loss due to radio frequency penetration and slow nuclear relaxation limits data collection. Since the materials are in a mixed superconducting state below $T_{\rm c}$, a macroscopic diamagnetic contribution $K_{\rm D}$ that is not precisely known appears. Since $K_{\rm D}$ is isotope independent for a given sample orientation, it can be removed by taking shift differences, e.g., ${^{63}K}_{\eta}-{^{199}K}_{\eta}$.    

\begin{figure}
\includegraphics[scale=0.75]{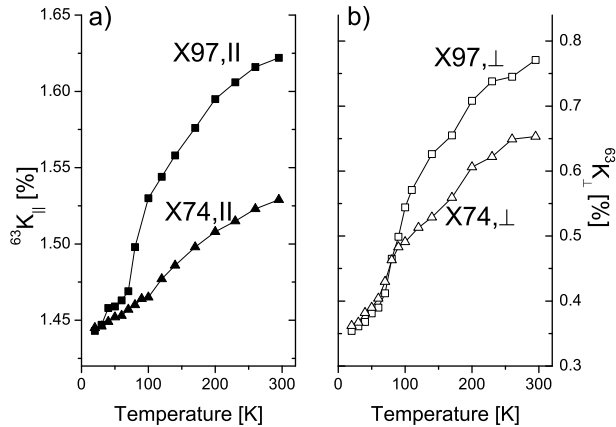}
\caption{$T$ dependencies of the \textit{total} magnetic shifts $^{63}K_{\parallel,\bot}(T)$ for both crystals, X74 ($T_{\rm c}$ = 74K ) and X97 ($T_{\rm c}$ = 97K); a) $\rm {c\parallel B_0}$, and b) $\rm {c\bot B_0}$. The quadrupole interaction was determined and its effect on the shift removed for $\rm {c\bot B_0}$.}
\label{fig:63KvsT}
\end{figure}

\section{Results and Discussion}
Fig.~\ref{fig:63KvsT} shows the $T$ dependence of the total magnetic Cu shifts $^{63}K_{\eta}$ (the error bars are within the size of the symbols) that are the sum of orbital, $K_{\rm orb}$, and spin, $K_{\rm S}$, contributions. Note that this cuprate is known to show a $T$ dependent Cu shift also for $\rm {c\parallel B_0}$ \cite{Suh1996,Itoh1998,Gippius1999}, in contrast to, e.g., YBa$_{\rm 2}$Cu$_{\rm 3}$O$_{\rm 7-\delta}$ and La$_{2-x}$Sr$_{x}$CuO$_4$. The term due to the partly diamagnetic response in the mixed state, $K_{\rm D}$, that we expect below $T_{\rm c}$, can be neglected for the large Cu shifts, as we show below when we discuss the $^{199}$Hg NMR data. Note that the anisotropies of the shifts (compare the actual numbers in both panels in Fig.~\ref{fig:63KvsT}) are due to the orbital  (Van Vleck) susceptibility. We show the total magnetic shifts (verified with field dependent measurements) to keep the discussion transparent. 

In order to isolate the spin shift one usually assumes that $K_{\rm orb}$ is largely $T$ independent (as are the quadrupole interaction and the hyperfine coefficients), and that the $T$ dependence of the shift is given by a $T$ dependent spin shift that vanishes below $T_{\rm c}$. This means $\chi_{{\rm S}}(T \rightarrow 0)\rightarrow 0$ due to spin singlet pairing, and thus the low-$T$ shifts must be given by the orbital term.

  The fact that the shifts for X97 and X74 at the lowest $T$ and given orientation are similar for each orientation in Fig.~\ref{fig:63KvsT} shows that the orbital shift term does not depend significantly on doping (similar to, e.g., La$_{2-x}$Sr$_{x}$CuO$_4$ \cite{Ohsugi1994}). At higher $T$ the $T$ dependence must then be due to the $T$ dependent spin susceptibility. 

\begin{figure}
\center{}
\includegraphics[scale=0.75]{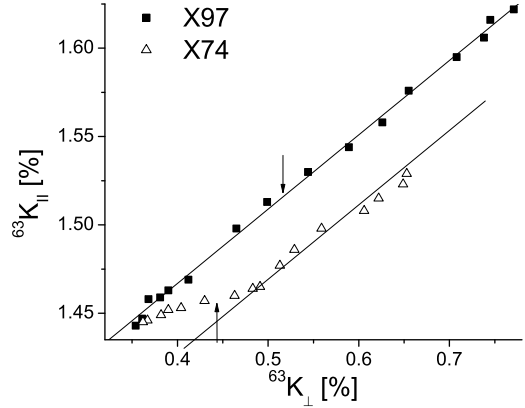}
\caption{ $^{63}K_\parallel(T)$ versus $^{63}K_\perp(T)$ with $T$ as an implicit parameter, for both single crystals, X74 and X97. The two arrows indicate $T_{\rm c}$ for both crystals. The two straight lines are linear fits to the data over the entire range for X97 and above $T_ {\rm c}$ for X74.}
\label{fig:63KK}
\end{figure}

Let us assume momentarily that the shifts in Fig.~\ref{fig:63KvsT} are caused by a single component spin susceptibility $\chi_{\rm S}(T)$. Then, all four curves in Fig.~\ref{fig:63KvsT} are described by $^{63}K_{{\rm S},\eta}(x_{j},T)={\rm ^{63}p}_{\eta} \cdot \chi_{\rm S}(x_{j},T)$ where $\eta$ denotes the orientation of the crystal with respect to the magnetic field and $x_{1}$, $x_2$ denote the doping. If we plot for a given sample the shifts measured for one orientation against that measured for the other orientation we expect straight lines with similar slopes. This can be seen by substituting $ \chi_{\rm S}(x_{j},T)$, e.g., ${^{63}K_{{\rm S},\parallel}(x_{j},T)}={\rm ^{63}p}_{\parallel}/ {\rm ^{63}p}_{\bot} \cdot {^{63}K_{{\rm S},\bot}(x_{j},T)}$.
Such plots are shown in Fig.~\ref{fig:63KK} (the actual data are the same as those presented in Fig.~\ref{fig:63KvsT}). We observe indeed straight lines with similar slopes at higher $T$ (larger shifts), however, there is an offset between them. Note that the similar slopes prove that the ratio of the hyperfine coefficients, and thus most likely the coefficients themselves, is independent on doping (the same conclusion is reached when comparing Cu and Hg data, see below). For X97 the straight line continues through $T_{\rm c}$ down to the lowest temperatures, while for X74 the measured data points move away from the straight line near $T_{\rm c}$ to meet those for the X97 material at the lowest $T$. That they meet at low $T$ is expected from Fig.~\ref{fig:63KvsT}, the orbital shifts are the same for both samples. Note that while we expect constant slopes, the constant high-$T$ offset between the two lines, which disappears at low $T$, cannot be explained. 

Clearly, the observation in Fig.~\ref{fig:63KK} that the changes of the X74 shifts with $T$ are not proportional to each other at all $T$ (here for different orientations of the field rather than different nuclei) demands a \textit{second} spin susceptibility with a different coupling to the nucleus. Thus, we are led to write\cite{Haase2009}
\begin{equation}
\label{eq:equ01}
K_{\eta}(x_j,T) ={\rm p}_{\eta} \cdot \chi_{1}(x_j,T) +  {\rm q}_{\eta} \cdot \chi_{2}(x_j,T),
\end{equation}
where $\chi_{2}(T)$ is the second $T$ dependent spin susceptibility and ${\rm q}_{\eta}$ the corresponding hyperfine coefficient. For a particular crystal ($x_j$) we then have for the data in Fig.~\ref{fig:63KK},
\begin{equation}
\label{eq:equ02}
K_{\parallel}(T)={{{{\rm{p}}_\parallel }} \over {{{\rm{p}}_\bot }}} \cdot K_{\bot}(T)+\left[{\rm q}_{\parallel}-{{{{\rm{p}}_\parallel }} \over {{{\rm{p}}_\bot }}}\cdot {{\rm{q}}_\bot}\right] \cdot \chi_{2}(T).
\end{equation}
Then, $\chi_{2}({\rm X97},T)$ must be rather small and $\chi_{2}({\rm X74},T>T_{\rm c})=const.$, but vanishes below about $T_{\rm c}$, and both sets of data in Fig.~\ref{fig:63KK} meet at low $T$. 

The need for two shift components is already obvious in Fig.~\ref{fig:63KvsT} where the shifts for both orientations of the X97 crystal are drawn to scale in panels a) and b). Then, since the shifts in each panel are the same at low $T$ the shifts for  both orientations of the X74 crystal (lower curves in both panels) should look very similar on this scale, as well. This is clearly not the case and Fig.~\ref{fig:63KvsT} tells us that the shifts cannot be understood with a single-component spin susceptibility. Instead, we need at least two components with different $T$ dependences. We can also conclude from the data in Fig.~\ref{fig:63KvsT} that both components have to disappear at low $T$. 

We would like to point out that the two susceptibilities that couple with ${\rm p}_\eta \ne {\rm q}_\eta$ to the nucleus, cf. (\ref{eq:equ01}), may not be the suceptibilities of the actual spin components. For example, a second spin component that couples to the nucleus through the same ${\rm p}_\eta$ as the first one could be present in the shift data for the X97 crystal in Fig.~\ref{fig:63KvsT} while not showing up in Fig.~\ref{fig:63KK}. However, we do infer from Fig.~\ref{fig:63KK} that $\chi_{2}$ is independent of $T$ above $T_{\rm c}$ for the X74 crystal. We refer to a more detailed discussion further below.

\begin{figure}
\center{}
\includegraphics[scale=0.7]{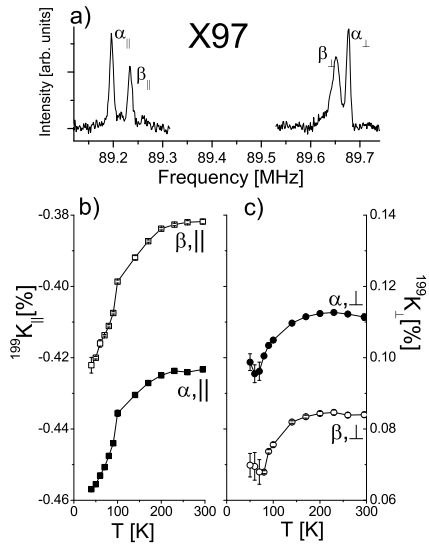}
\caption{$^{199}$Hg NMR of the $T_{\rm c}$ = 97K single crystal (X97); a), typical spectra for both directions of the field show two distinct Hg sites; b) and c), $T$ dependencies of the shifts for the lines shown above for $\rm {c\parallel B_0}$ and $\rm {c\bot B_0}$, respectively.}
\label{fig:199K}
\end{figure}

\

We now turn to $^{199}$Hg NMR. We show in Fig.~\ref{fig:199K} data for the X97 crystal: a pair of typical spectra for both orientations of the field (a), and the $T$ dependencies of the corresponding shifts (b, c). We observe two well resolved $^{199}$Hg lines and label them with $\alpha$ and $\beta$. Note that the $T$ dependence of the Hg shifts is not very different from that of Cu, but the shifts are smaller, as one expects from transferred hyperfine coupling to the more distant Hg atoms. Similar results were obtained for the X74 crystal (data not shown), which reveal that the $\alpha$ line is the same as for the X97 crystal at the lowest $T$ where the spin shift is small, whereas the $\beta$ line has a smaller shift and lower intensity. Therefore, the $\beta$ Hg site must be due to Hg atoms that are affected by doping (for further details we refer to \cite{Rybicki2011a}). Here we would like to focus on the X97 crystal in order to test the single vs. two component description for this system.

\begin{figure}
\center{}
\includegraphics[scale=0.85]{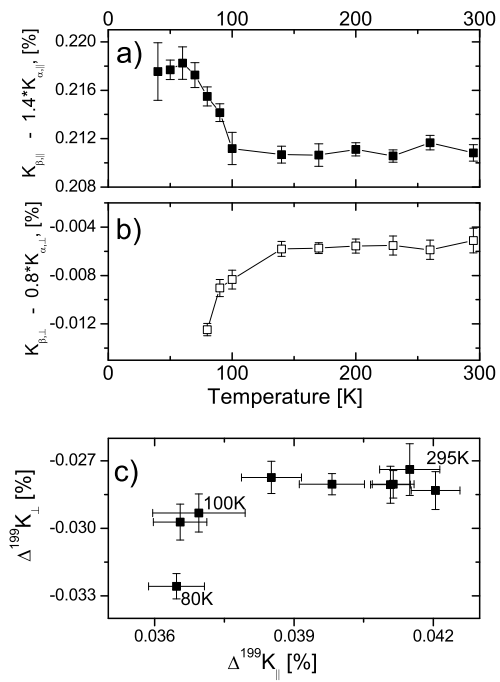}
\caption{X97 crystal: a) , b) $T$ dependence of ${^{199}K_{\beta,\parallel}} -1.4\cdot  {^{199}K_{\alpha,\parallel}}$ and ${^{199}K_{\beta,\bot}} -0.8\cdot  {^{199}K_{\alpha,\bot}}$, respectively. c) $\Delta {^{199}K_{\bot}} \equiv {^{199}K_{\beta,\bot}} - {^{199}K_{\alpha,\bot}}$ vs. \mbox{${\Delta ^{199}K_{\parallel}} \equiv {^{199}K_{\beta,\parallel}} - {^{199}K_{\alpha,\parallel}}$} with $T$ as an implicit parameter (see text).}
\label{fig:K97vsK74}
\end{figure}

Since we have two sharp, well resolved Hg resonances we can plot their shifts against each other, similar to what we did for $^{63}$Cu, to inquire about the spin susceptibility. Note, however, that the orbital contributions to the shifts are very large compared to the $T$ dependent shifts, see both panels in Fig.~\ref{fig:199K}. It is therefore most reliable to plot the shifts of both lines $\alpha$ and $\beta$ for each orientation $\eta$ of the field against each other, i.e., $^{199}K_{\beta,\eta}$ vs. $^{199}K_{\alpha,\eta}$. From such plots (see below) we find straight lines with fixed slopes for $T>T_{\rm c}$. Since $K_{\rm D}$=0 for $T>T_{\rm c}$ we infer that there is only one $T$ dependent spin susceptibility for $T>T_{\rm c}$. We find for the slopes ${\rm r_\parallel} = 1.4$ and ${\rm r_\bot} = 0.8$ for $\rm {c\parallel B_0}$ and $\rm {c\bot B_0}$, respectively. In order to demonstrate this with higher graphical resolution we plot in Fig.~\ref{fig:K97vsK74} (a, b),  $\left[{{^{199}K_{\beta,\eta}}(T) - {\rm r}_\eta \cdot  {^{199}K_{\alpha,\eta}}(T) }\right]$ vs. $T$. We can look at (\ref{eq:equ02}) to discuss Fig.~\ref{fig:K97vsK74} (a, b).  Since ${\rm r}_\eta = {^{199}{\rm p}_{\beta,\eta}} / {^{199}{\rm p}_{\alpha,\eta}}$, horizontal lines for $T>T_{\rm c}$ in Fig.~\ref{fig:K97vsK74} (a, b) are equivalent, cf. (\ref{eq:equ02}), to constant slopes ${\rm r}_\eta$ above $T_{\rm c}$ in the corresponding shift-shift plots. Below $T_{\rm c}$ we observe deviations from the horizontal lines for both orientations, a clear indication of another shift contribution. However, we cannot conclude from it with (\ref{eq:equ02}) on a second $T$ dependent spin component since $K_{\rm D}$ could cause such a behavior, as well.

In order to remove the uncertainty due to $K_{\rm D}$ we now take the differences between the shifts of the two lines for each orientation. With $\Delta {^{199}K_{\eta}} \equiv {^{199}K_{\beta,\eta}} - {^{199}K_{\alpha,\eta}}$ we have with (\ref{eq:equ01}), here adopted for the $^{199}$Hg NMR spin shift,
\begin{equation}
\label{eq:equ05}
\Delta {^{199}K_{\eta}}=\Delta{\rm ^{199}p}_{\eta} \cdot \chi_{1}(T) +  \Delta{\rm ^{199}q}_{\eta} \cdot \chi_{2}(T),
\end{equation}
where $\Delta{\rm ^{199}p}_{\eta}\equiv {\rm ^{199}p}_{\beta, \eta}-{\rm ^{199}p}_{\alpha, \eta}$, etc. 
Analogous to (\ref{eq:equ02}) we can express one shift difference in terms of the other in order to remove one of the susceptibilities, and we have (dropping the mass number of the isotope for convenience)

\begin{equation}
\label{eq:equ06}
\Delta {K_ \bot } = {{\Delta {{\rm{p}}_ \bot }} \over {\Delta {{\rm{p}}_\parallel }}} \cdot \Delta {K_\parallel } + \left[ {\Delta {{\rm{q}}_ \bot } - {{\Delta {{\rm{p}}_ \bot }} \over {\Delta {{\rm{p}}_\parallel }}} \cdot \Delta {{\rm{q}}_\parallel }} \right] \cdot {\chi _2}.
\end{equation}

This is shown in Fig.~\ref{fig:K97vsK74} (c), however, there we plot the difference of the \textit{total experimental shifts} between the $\beta$ and $\alpha$ line for $\rm {c\bot B_0}$ as a function of the same difference for $\rm {c\parallel B_0}$. Note that $T$ independent orbital terms cause a constant offset on both axes.
For a single $T$ dependent spin component, i.e., $\chi_2=0$ in (\ref{eq:equ06}), a straight line with a fixed slope is expected. This is clearly not the case. We observe that the shift difference for $\rm {c\parallel B_0}$ drops significantly already at high $T$ as the $T$ is lowered, while that for $\rm {c\bot B_0}$ is constant at high $T$, but drops only below about $T_{\rm c}$, a behavior that cannot be explained by a single $T$ dependent spin susceptibility. From (\ref{eq:equ06}) we conclude that we need a second susceptibility that is $T$ independent above $T_{\rm c}$, but changes below $T_{\rm c}$. This independent experimental observation proves that a second component is necessary also for the X97 crystal even if it appears to be small. While this is in agreement with what we derived from comparing the Cu shifts for X97 and X74, it was not obvious from the Cu shifts for the X97 crystal alone.

\

\begin{figure}
\center{}
\includegraphics[scale=0.7]{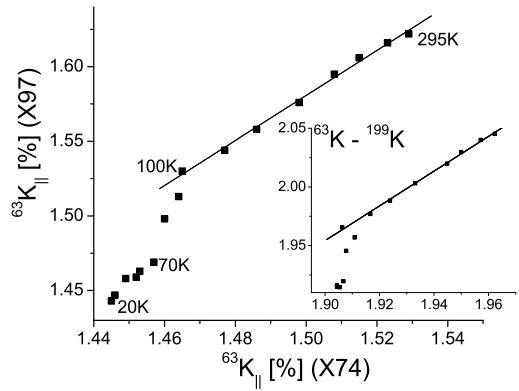}
\caption{Main panel: ${^{63}K_{\parallel, \rm {X97}}(T)}$  vs.  ${^{63}K_{\parallel,\rm {X74}}(T)}$. In order to remove a possible diamagnetic term $K_{\rm D}$ the Hg shifts have been subtracted from the Cu shifts in the inset: $\left[{^{63}K_{\parallel, \rm {X97}}(T)}-{^{199}K_{\parallel,\alpha,\rm {X97}}(T)}\right]$  vs.  $\left[{^{63}K_{\parallel,\rm {X74}}(T)}-{^{199}K_{\parallel,\alpha, \rm {X74}}(T)}\right]$.}
\label{fig:X97vsX74}
\end{figure}

Somewhat unusual plots are presented in Fig.~\ref{fig:X97vsX74}. In the main panel we see the Cu shift with ${\rm c \parallel B_0}$ for one crystal (X97, vertical axis) plotted against the Cu shift with ${\rm c \parallel B_0}$ of the other crystal (X74, horizontal axis). Surprisingly, one finds a straight line above the X97 $T_{\rm c}$. This says that, coming from high $T$ (large shifts), the shifts in Fig.~\ref{fig:63KvsT} for both crystals fall in such a way that leaves the ratio of the slopes constant. At the $T_{\rm c}$ of the X97 crystal its shift drops rapidly and below 70 K we find a behavior similar to the one observed at higher $T$. The fact that they arrive at similar shift values for the lowest $T$ has been noted in Fig.~\ref{fig:63KvsT} and is expected for a doping independent orbital shift. 
Since we know that $\chi_2(T>T_{\rm c})=const.$ we conclude from Fig.~\ref{fig:X97vsX74} for $T>T_{\rm c}$ that since  $\left[\Delta K_{\rm X97}\right]/\left[\Delta K_{\rm X74}\right] \approx 1.4$ we must have  $\Delta\chi_{1}({\rm X97})/\Delta \chi_{1}({\rm X74}) \approx 1.4$ for all $T$ intervals. Thus, $\chi_{1}({\rm X97})=1.4\cdot \chi_{1}({\rm X74})+c$, where $c$ is a constant that must have vanished below 70K in Fig.~\ref{fig:X97vsX74}. This says that $\chi_{1}$ consists of a component that is $T$ dependent at all $T$ and changes proportionally with doping, plus a second component with a behavior similar to $\chi_2$.

Before we continue with the discussion we would like to prove that the diamagnetic shift $K_{\rm D}$ is indeed not important when comparing Cu shifts. We show in the inset of Fig.~\ref{fig:X97vsX74} a plot similar to that in the main panel, but with the Hg shifts subtracted from the corresponding Cu shifts so that $K_{\rm D}$ is eliminated on both axes. As expected, we observe again a straight line above $T_{\rm c}$ and since the Hg shifts are small it has a similar slope. The drop just below $T_{\rm c}$ remains and this rules out the importance of $K_{\rm D}$ on the Cu shift scale.

\begin{figure}
\center{}
\includegraphics[scale=0.7]{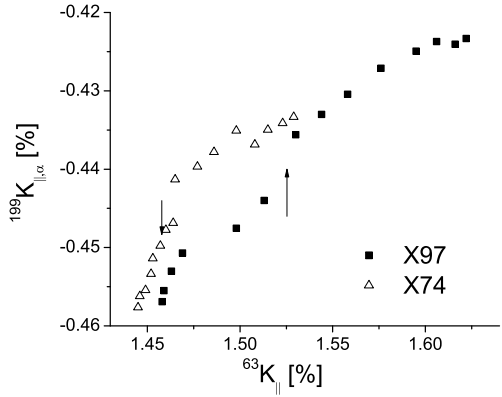}
\caption{X74 and X97 crystals: $^{199}K_{\parallel,\alpha}$ vs. $^{63}K_{\parallel}$ for both crystals with $T$ as an implicit parameter (see text).}
\label{fig:199Kvs63K}
\end{figure}

We plot in Fig.~\ref{fig:199Kvs63K} the Hg shifts (total experimental shifts) of the $\alpha$ lines against those of Cu for both crystals and {$c\parallel {\rm B}_0$}. As expected, the slopes at higher $T$ are approximately straight lines and similar (note that larger errors are expected since the Cu shifts are much bigger than those of Hg). We find that ${{^{199}{\rm p}_{\parallel, \alpha}}/ {^{63}{\rm p}_{\parallel}} \approx 0.12}$. For the X97 crystal we observe a steeper drop near $T_{\rm c}$, which levels off to the slope above $T_{\rm c}$ before it finally drops much more rapidly with lowering $T$. Since the first drop below $T_{\rm c}$ for the X97 crystal is accompanied by a large shift change for Cu, it is not dominated by $K_{\rm D}$. The second and steep drop for both crystals at low $T$ must be mostly due to $K_{\rm D}$ since its slope is close to 1 ($K_{\rm D}$ is independent of the nucleus). With this we can estimate the maximum diamagnetic shift for ${\rm c \parallel B_0}$ to be $\Delta K_{\rm D}< 0.015\%$. Thus, the contributions to the $T$ dependent shift below $T_{\rm c}$ from  $K_{\rm D}$ and  $K_{\rm S}$ are of similar size for $^{199}$Hg (but not for $^{63}$Cu).
\\

Since we need two spin component (we call them A and B) for the explanation of our data, we would like to point out that this demands in general\cite{Curro2004} \textit{three} spin susceptibilities\cite{Haase2009}: $\chi_{\rm AA}, \chi_{\rm BB}$, and $\chi_{\rm AB}$ since there can be a coupling of the two spin components A and B. Accordingly, we have ${\chi_1=\chi_{\rm AA}+\chi_{\rm AB}}$, and $\chi_2=\chi_{\rm AB}+\chi_{\rm BB}$. The total spin susceptibility is then given by $\chi_{\rm S}=\chi_{\rm AA}+2\chi_{\rm AB}+\chi_{\rm BB}$. For the NMR spin shifts it follows, 
\begin{equation}
\label{eq:equ07}
K_{\eta,X} = {\rm p}_{\eta} \left(\chi_{{\rm AA},X} +\chi_{{\rm AB},X}\right)+  {\rm q}_{\eta} \left(\chi_{{\rm BB},X}+\chi_{{\rm AB},X}\right). 
\end{equation}
We know that $\left(\chi_{{\rm BB},X}+\chi_{{\rm AB},X}\right)$ must be $T$ independent at higher $T$. We now assume that both component $\chi_{{\rm BB},X}$ and $\chi_{{\rm AB},X}$ indeed contribute to the shifts. Since it is highly unlikely that both components are $T$ dependent, but their sum is not, we conclude that both must be $T$ independent above $T_{\rm c}$. Then, $\chi_{\rm AA}$ must be the susceptibility that is $T$ dependent already above $T_{\rm c}$. 

We deduce from Fig.~\ref{fig:63KK} that ${\rm p}_\parallel$/${\rm p}_\bot \approx 0.40$ and with
\begin{equation}
\label{eq:equ09}
{C_{X}\equiv \left[{\rm q}_\parallel - ({\rm p}_\parallel/{\rm p}_\bot) \cdot {\rm q}_\bot \right] \left[\chi_{{\rm AB},X}+\chi_{{\rm BB},X}\right]}
\end{equation}
that  $C_{\rm X97} \approx 0\%$, $C_{\rm X74}({T>T_{\rm c}}) \approx -0.040\%$, and ${C_{\rm X74}(T \ll  T_{\rm c}) \approx~0\%}$. Thus, at high $T$ $\left[\chi_{{\rm AB, X74}}+\chi_{{\rm BB, X74}}\right] \approx const.$, but it vanishes rapidly below $T_{\rm c}$.  We also find  $\chi_{{\rm AB, X97}} \approx - \chi_{{\rm BB, X97}}$, and if one of the components is sizable this relation holds approximately also below $T_{\rm c}$. We now understand why the influence of the second spin component is rather weak in the plots in Figs.~\ref{fig:63KK} and \ref{fig:K97vsK74} (c) for X97. 

Note, that we discovered with Fig.~\ref{fig:X97vsX74} that $\Delta K_{\rm X97}/\Delta K_{\rm X74} \approx 1.4$ at higher $T$. A simple explanation follows by assuming that $\chi_{\rm AA, X97}\approx 1.4\cdot \chi_{\rm AA,X74}$ and that the integration constant is just due to $\chi_{{\rm AB},X}$. We then have with (\ref{eq:equ07}) for the curve in the main panel,
\begin{eqnarray}
\label{eq:equ06c}
\begin{split}
{^{63}{K_{\eta ,{\rm{X97}}}}} &= {r_\eta }^{}{}^{63}{K_{\eta ,{\rm{X74}}}} \\
&+ {}^{63}{\rm p_\eta } \cdot {\chi _{{\rm{AB}},{\rm{X97}}}} \\
 & {}+ {}^{63}{\rm q_\eta }\cdot \left[ {{\chi _{{\rm{AB}},{\rm{X97}}}} + {\chi _{{\rm{BB}},{\rm{X97}}}}} \right] \\
 & {}- r_\eta ^{}\left\{ {\left( {{}^{63}{\rm p_\eta } + {}^{63}{\rm q_\eta }} \right){\chi _{{\rm{AB}},{\rm{X74}}}} + {}^{63}{\rm q_\eta }\cdot{\chi _{{\rm{BB}},{\rm{X74}}}}} \right\}.
\end{split}
\end{eqnarray}
We note that the terms in the last row stem from the X74 cyrstal and will not contribute to the drop in Fig.~\ref{fig:X97vsX74} near $T_{\rm c} \approx 97$K. In addition, since the X74 shift for ${c\parallel B_0}$ in Fig.~\ref{fig:63KvsT} does not change significantly below $T_{\rm c}$ we know that the term in the last row can be ignored at lower $T$, as well. Similarly, the term in the third row was found to be rather small for the X97 crystal ($\chi_{\rm 2,X97}\approx 0$). We are left with the term in the second row, and since $\chi_{\rm AB}$ vanishes rapidly at $T_{\rm c} \approx 97$K the observed drop in Fig.~\ref{fig:X97vsX74} is expected. We conclude that this drop is mainly given by $^{63}{\rm p_\parallel } \cdot {\Delta \chi _{{\rm{AB}},{\rm{X97}}}}\approx +0.05\%$.

We can also inspect Fig.~\ref{fig:199Kvs63K} again. For the X97 crystal the change of the slope at $T_{\rm c}$ marks the change in $\chi_{\rm AB}$ and $\chi_{\rm BB}$ until $K_{\rm D}$ dominates at lower $T$. On the horizontal axes this is given by $\Delta {^{63}{K_{\parallel ,{\rm{X97}}}}}={^{63}{\rm p}_\parallel} \cdot \Delta \chi_{\rm AB, X97} \approx +0.05\%$, as concluded from Fig.~\ref{fig:X97vsX74}. The corresponding change in shift for Hg of about +0.01\% is somewhat bigger due to $K_{\rm D}$ (${^{199}{\rm p}_\parallel}/{^{63}{\rm p}_\parallel} \approx 0.12$). For the X74 crystal we do not expect a drop near $T_{\rm c}$ for Cu (cf. discussion above), and for Hg it is mostly given by $K_{\rm D}$. (It appears that the onset of the change in slope is slightly above $T_{\rm c}$ outside the estimated error in the $T$ measurement).

On general grounds we expect $\chi_{\rm AA}$ to be positive and therefore it follows that $^{63}{\rm p_\eta }>0$ for both orientations. Then, we have for $T>T_{\rm c}$ that $\chi_{\rm AB,X97}>0$, and since ${{\chi _{{\rm{AB}},{\rm{X97}}}} \approx -{\chi _{{\rm{BB}},{\rm{X97}}}}}$, it follow that $\chi_{\rm BB,X97}<0$ above $T_{\rm c}$ for this material. For the X74 cyrstal we notice, e.g., from Fig.~\ref{fig:63KvsT}, that the constant shift term above $T_{\rm c}$ must be small for $c\parallel {\rm B}_0$, but larger for $c\bot {\rm B}_0$.
\\

To conclude, based on a thorough NMR study of the two HgBa$_{2}$CuO$_{4+\delta}$ crystals \cite{Rybicki2009,Rybicki2011a} we identified a 2nd shift component that is independent of $T$ above $T_{\rm c}$ and vanishes rapidly below $T_{\rm c}$. Since it influences the Cu and Hg nuclei by similar relative values, the 2nd component must be due to a spin susceptibility, in accord with what some of us proposed earlier for La$_{1.85}$Sr$_{0.15}$CuO$_4$ \cite{Haase2009}, and very recently for YBa$_{\rm 2}$Cu$_{\rm 4}$O$_{\rm 8}$ \cite{Meissner2011}. Our results can be  explained consistently with two spin components that are coupled to each other and thus have the three susceptibilities $\chi_{\rm AA}, \chi_{\rm AB}, \chi_{\rm BB}$. In such a scenario, cf. (\ref{eq:equ07}), we find that only one of them, $\chi_{\rm AA}$ is $T$ dependent above $T_{\rm c}$, while the other two vanish below $T_{\rm c}$. Furthermore, we find that $\chi_{\rm AA,X97}=1.4 \cdot \chi_{\rm AA,X74}$, a relation that has been suggested by some of us earlier\cite{Rybicki2009} and that some of us applied successfully to understand the changes in the $^{17}$O NMR spin shift of YBa$_{\rm 2}$Cu$_{\rm 4}$O$_{\rm 8}$ with pressure\cite{Meissner2011}. Thus, $\chi_{\rm AA}(T)$ is responsible for the pseudogap feature observed in NMR and it changes proportionally (increases) with increasing doping for an underdoped cuprate, but decreases with increasing pressure \cite{Meissner2011}.  (We believe that $\chi_{\rm AA}(x_2)/\chi_{\rm AA}(x_1)= x_2/x_1 $ where $x_j$ is the local hole doping per CuO$_2$ in the plane, so that $\chi_{\rm AA}=\rho(T)\cdot x$, see \cite{Rybicki2009}). The part of the shift that is constant above $T_{\rm c}$ was found to increase dramatically with pressure\cite{Meissner2011} and we believe that it mus also increase with doping. However, more experiments are necessary to explore the two components across the phase diagram and as a function of pressure. Nevertheless, there can be no doubt that a two-component scenario is generic to the cuprates. This also means that the hitherto adopted single-component interpretation of NMR data can no longer be upheld.

We can only guess the physical origin of the two spin components. Barzykin and Pines in an extensive review \cite{Barzykin2009} have put forward a model. Although the $T$ dependence of the first component reminds one of coupled Heisenberg spins \cite{Johnston1989}, we note that the pseudogap regime of the cuprates has recently been shown to be associated with unconventional magnetism \cite{Li2008,Li2010}. Whereas the 2nd component exhibits the behavior of a Fermi liquid that becomes superconducting at $T_{\rm c}$, a $T$ constant behavior at higher $T$ is also expected for a Fermi glass \cite{Mueller1981}.

\section{Acknowledgement}
We acknowledge the help of M. Lux, M. Jurkutat, T. Meissner, M. Bertmer, A. P\"oppl; the financial support by the University of Leipzig, and discussions with O.P. Sushkov (JH, DR). The crystal growth work was supported by the Department of Energy under Contract No. DE-AC02-76SF00515
and by the National Science Foundation under Grant No. DMR-0705086.


\end{document}